\documentclass{article}

\newcommand{\vx}{{\bf x}}
\newcommand{\vk}{{\bf k}}
\newcommand{\mR}{{\bf R}}

\usepackage{arxiv}
\usepackage{amssymb}
\usepackage{amsmath}

\usepackage{graphicx}
\usepackage{float}
\restylefloat{figure}

\usepackage{biblatex}
\usepackage[utf8]{inputenc} 
\usepackage[T1]{fontenc}    
\usepackage{hyperref}       
\usepackage{url}            
\usepackage{booktabs}       
\usepackage{amsfonts}       
\usepackage{nicefrac}       
\usepackage{microtype}      

\title{Single-Pixel Fluorescent Diffraction Tomography}

\author{
  Patrick A. Stockton \\
  Department of Electrical and Computer Engineering\\
  Colorado State University\\
  Ft. Collins, CO 80528 \\
   \And
 Jeffery J. Field \\
  Department of Electrical and Computer Engineering\\
  Colorado State University\\
  Ft. Collins, CO 80528 \\
   \And
  Jeff Squier\\
  Department of Physics\\ 
  Colorado School of Mines\\
  Golden, CO 80401\\
   \And
  Ali Pezeshki\\
   Department of Electrical and Computer Engineering\\
   Colorado State University\\
   Ft. Collins, CO 80528 \\
   \AND
   Randy A. Bartels\\
   Department of Electrical and Computer Engineering\\
   Colorado State University\\
   Ft. Collins, CO 80528 \\
   \texttt{randy.bartels@colostate.edu} \\
}

\begin{document}

\maketitle

\begin{abstract}
Optical diffraction tomography is an indispensable tool for studying objects in three-dimensions due to its ability to accurately reconstruct scattering objects. Until now this technique has been limited to coherent light because  spatial phase information is required to solve the inverse scattering problem. We introduce a method that extends optical diffraction tomography to imaging spatially incoherent contrast mechanisms such as fluorescent emission. Our strategy mimics the coherent scattering process with two spatially coherent illumination beams. The interferometric illumination pattern encodes spatial phase in temporal variations of the fluorescent emission, thereby allowing incoherent fluorescent emission to mimic the behavior of coherent illumination. The temporal variations permit recovery of the propagation phase, and thus the spatial distribution of incoherent fluorescent emission can be recovered with an inverse scattering model. 
\end{abstract}


\section{Introduction}
In fluorescence microscopy, light emitted from the specimen is spatially incoherent. Consequently, 3D imaging techniques require some form of spatial gating to map detected photons to the location from which they were emitted. This spatial gating is often achieved though some combination of confining the illumination volume and detection volumes. Examples of such strategies include selective plane illumination microscopy (SPIM) \cite{spim}, where the illumination volume is restricted to a thin axial plane, or laser-scanning confocal microscopy, where both the illumination and detection volumes are restricted to a diffraction-limited spot in 3D \cite{confocalTony}. These strategies allow each detected photon to be mapped to a 3D location in the specimen from which it was emitted, and often involve high numerical aperture (NA) optics that tightly focus the illumination light and/or restrict the volume over which light is detected. 

Other strategies for 3D imaging rely on inverting a quantitative model of the illumination, emitted, and collected light to estimate the concentration of fluorescent emitters from detected intensity images. These computational imaging methods, such as optical projection tomography (OPT), require axially scanning or rotating the object to collect a set of data to be reconstructed \cite{Comparison2019}. Conventional fluorescent imaging methods suffer from limitations such as photobleaching and anisotropic spatial resolution between the axial and transverse directions \cite{confocal}. While SPIM can partially mitigate the effects of photobleaching, anisotropic spatial resolution is a persistent problem \cite{Comparison2019}. In all of these methods, tissues must be optically cleared to reduce distortions from optical scattering to suitably low levels \cite{Comparison2019}. A more stringent restriction on SPIM and OPT microscopes is that the spatial resolution is coupled to the size of the object \cite{Comparison2019} -- leading to decreased spatial resolution for in increased imaging region. 

\begin{figure}[t!]
      \centering
      \fbox{\includegraphics[width=0.9\linewidth]{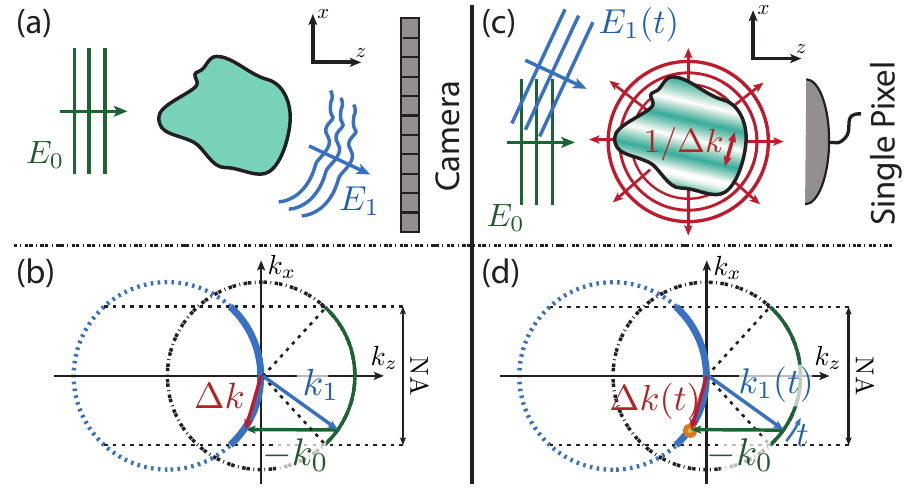}}
      \caption{ODT compared to FDT. (a) The standard optical diffraction tomography scattering picture. (b) The spatial frequency support probed by diffraction tomography. The thick blue arc is the measured spatial frequency information. (c) The FDT picture, with the signal light collected by a single-pixel detector. (d) The spatial frequency representation of FDT.}
      \label{fig:Scattering}
        \vspace{0.5\baselineskip}
    \end{figure}

Coherent imaging strategies enable 3D imaging by making use of the direction of the scattered light. Emil Wolf recognized that the inverse scattering problem for coherent light propagating in an object can be solved by recording the complex, spatially coherent, scattered field \cite{WOLF1969}. Directional scattering allows the recording of spatial frequency components of an object by exploiting knowledge of the complex amplitude of light scattered in a particular direction when illuminated by a spatially coherent input wave. This concept is illustrated schematically in Fig. \ref{fig:Scattering}(a), where the illumination field, $E_0$, and scattered field, $E_1$, have corresponding wavevectors ${\bf k}_0$ and ${\bf k}_1$ respectively. 

Interferometric techniques, e.g., holography, are able to record the complex scattered field that, within the Born approximation, can be mapped to an arc of spatial frequency information defined by the Ewald sphere by applying the Fourier diffraction theorem \cite{WOLF1969} shown in Fig. \ref{fig:Scattering}(b). The position in spatial frequency space is given by the wavevector difference, ${\bf \Delta k} = {\bf k}_1 - {\bf k}_0$. This sparse spatial frequency information is encoded by the complex scattered field obtained in a single scattered field measurement. More complete object information can be acquired by introducing a relative rotation between the illumination and the object to fully sample the object spatial frequency distribution, yielding optical diffraction tomography (ODT). 

Optical holography, and thus ODT, normally relies on spatially coherent light to interferometrically record the complex scattered field, allowing interior spatial frequency information to be acquired. Coherent scattering data can be inverted to solve the scattering problem for variations in refractive index of the specimen. Using coherent illumination allows object position to be encoded in the complex scattered field. The phase is critical since it encodes the axial location of the scatterer, and it is this phase that is required to enable diffraction tomography to be extended to incoherent fluorescent light. ODT uses a rotation of the object or illumination wave to capture a sequence of scattered fields that fill out the object spatial frequency information. Then computational imaging tools are applied to invert the information recorded in order to recover the object spatial frequency distribution, and thus the object spatial information. ODT has the advantage of not being constrained to imaging objects in the Rayleigh range of the illumination beam, as the light is allowed to diffract before encountering the optical detector.  

Such a coherent imaging method is conventionally thought to be impossible with fluorescent light because, in the case of incoherent emission, phase is lost due to the random emission of the molecular emitters and this unstable phase obscures the relationship between the location of the emitter and the propagation direction and phase. If fluorescence is to be imaged in a similar manner as ODT, it is necessary to encode the coherent illumination propagation phase onto fluorescent light so that the phase can be recovered. It is possible to exploit the fact that an incoherent emitter is coherent with itself to encode spatial location of the emitter \cite{FINCH2016, SELFI2019}, but a general adaptation of coherent-like imaging methods to incoherent light remains elusive. 

 In this Letter, we introduce the first method that uses fluorescent light emission for diffraction tomographic imaging. As fluorescent light is spatially incoherent, it is necessary to mimic the process of coherent scattering to enable optical diffraction tomography with fluorescence. We mimic spatially coherent scattering by transferring the phase difference from a pair of spatially coherent illumination beams \cite{Field:16} into the a time-variation of fluorescent emission brightness, enabling ODT with fluorescent light recorded on a single element optical detector. By mimicking the incident and scattered fields in the illumination of a fluorescent object, we are able to perform optical diffraction tomography using fluorescent light. We refer to this method as Fluorescence Diffraction Tomography (FDT).
 
The FDT concept is illustrated in Fig. \ref{fig:Scattering}(c) and (d). A pair of illumination beams substitute for the incident and scattered waves in coherent scattering. The reference wave, $E_0$, in Fig. \ref{fig:Scattering}(c), plays the role of the incident wave in coherent scattering and interferes with an illumination plane wave, $E_1$, that represents the scattered wave. To map out the equivalent information as in coherent scattering, the incident direction of $E_1$ is scanned in time, producing a modulation of the illumination intensity that depends on the relative phase of the two illumination beams, $\Delta {\bf k}(t) \cdot \vx$, where ${\vx}=(x,z)$ is the spatial coordinate vector in the $x-z$ plane. The difference wavevector, $\Delta {\bf k}(t) = {\bf k}_1(t) - {\bf k}_0$, behaves as the scattering vector in coherent scattering that is defined at the difference between the k-vector of the scattered field, ${\bf k}_1$, and incident waves, ${\bf k}_0=k(0,1)$, where $k=2\pi/\lambda$ is the wavenumber of the illumination.
 
The collected fluorescence  recorded with a single-pixel detector serves as the FDT time signal. These measurements imprint the relative phase of the two spatially coherent illumination beams into an intensity modulation in space and time that allows the detected incoherent fluorescent light power to be treated as if it came from a coherent source. Because fluorescent light is incoherent, the detected fluorescent light power is equivalent to the overlap integral between the spatial distribution of the fluorophore concentration -- our object -- and the illumination intensity. Each measurement at time $t$ samples the complex amplitude of the object spatial frequency distribution at the difference spatial frequency wavevector, $\Delta {\bf k}(t)$.
The result is that for each incident angle of $E_1(t)$, a spatial frequency projection is recorded that exactly mimics the complex spatial frequency information traditionally obtained through coherent scattering measurements; compare Figs. \ref{fig:Scattering}(b) and (d).  

The key aspect of FDT is coherent transfer mediated by the modulated illumination intensity. The intensity arises from the interference of the reference and scanned fields, i.e., $I_{\rm ill} \propto \big|E_0+E_1[\theta(t)]\big|^2$ \cite{Field:16}, where $\theta(t)$ denotes the incidence angle of $E_1$ with respect to $k_z$ axis, or equivalently the angle between $\vk_1$ and $\vk_0$, at time $t$. The model for the interference is written as $I_{ill}(\vx,t) = 1 + \mu(t) \cos[\Delta \Phi(\vx,t)]$, where $\mu(t)$ is the fringe visibility and $\Delta \Phi$ is the phase difference between the reference and scanned fields in Fig. \ref{fig:Scattering}(c) \cite{Field2015, Field:2019}. The phase difference between the illumination fields, $\Delta \Phi({\vx},t) = \omega_c \, t + \Delta {\bf k}(t) \cdot \vx$, imparts a temporal modulation pattern at each spatial position in the $x-z$ plane. Here, $\omega_c$ is a carrier frequency in the modulation, which is critical for isolating the complex phase information in the time signal \cite{Field2015, Field:16}, and $\Delta {\bf k}(t) \cdot \vx$ is the spatial phase variation that encodes the location of the object in the $x-z$ plane. 

The elements of $\Delta \vk(t)= (\Delta k_x(t),\Delta k_z(t))$ are difference frequencies. In our work, we set $\Delta k_x(t)=k_c \, t/T$, corresponding to  $\sin\theta(t)= t \, {\rm NA}/T$,  where $k_c=k \, {\rm NA}$ is the coherent imaging cutoff spatial frequency for the illumination optics and $t\in [-T,T]$ with $2 T$ denoting the total collection time. For this choice, we have $\Delta k_z=k \, (\sqrt{1-({t\, \rm NA}/T)^2}-1)$. The spatio-temporal intensity modulation encodes the relative spatial phase of the illumination fields as a temporal modulation of the emitted fluorescent power from the object, thereby transferring coherent propagation behavior to fluorescent emission \cite{Field2015}.  

The time trace is generated by detecting the collected fluorescent emission as the scanning field sweeps through the range of incident angles supported by the NA of the illumination objective \cite{Field2015,Field:16,Field:2019}. The temporal signal $S(t,\phi)$ is the projection of the spatial distribution $c(\vx)$ of the fluorophore concentration onto illumination intensity at incidence angle $\phi$:
    \begin{equation}
        S(t,\phi) = \langle I_{ill}(\mR_\phi \vx,t) c(\vx) \rangle_{\vx}
        \label{eqn:sigmod}
    \end{equation}
where Dirac integral notation, $\langle \cdot \rangle_{\vx} = \int \cdot\ \mathrm{d}\vx$, denotes the spatial integration over $x$ and $z$, performed by the single-pixel detector, and $\mR_\phi$ is a rotation matrix by $\phi$ that yields the coordinate transform  $\vx=(x,z) \longrightarrow \mR_{\phi}\vx=(x \cos\phi - z \sin\phi,x \sin\phi + z \cos\phi)$. We have not included a measurement noise term $\varepsilon(t)$ in writing Eq. (\ref{eqn:sigmod}) to keep subsequent equations simple in form, but it is understood that an additive noise term is always present.

    \begin{figure}[tb!]
        \centering
        \fbox{\includegraphics[width=0.9\linewidth]{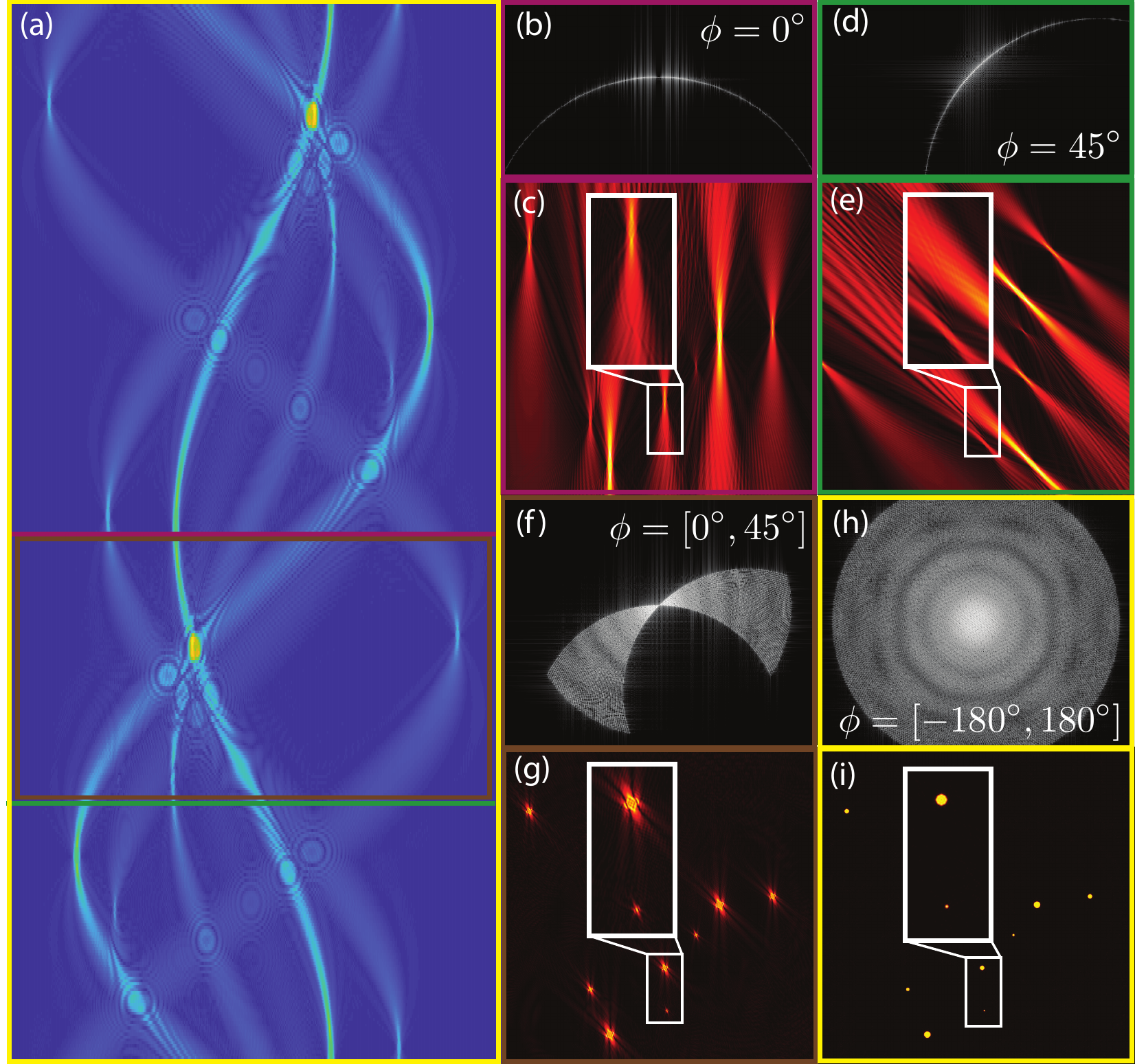}}
      {\caption{FDT sinogram and reconstruction. (a) Simulated FDT sinogram from a fluorescence distribution where the object has been rotated over $[-180^{\circ},180^{\circ}]$. (b), (d), (f), and (h) show the frequency support that is measured and mapped onto the Ewald sphere and $\phi$ indicates the illumination angle in degrees.  (c), (e), (g), and (i) show the reconstructed object generated from by applying the dual operator to the FDT time signal, Eq. (\ref{eqn:inverse_Model}). The colored boxes around panels (b-i) correspond to the colored lines and boxes in panel (a) and represent the measured information used in the reconstruction.}
      \label{fig:reconstruction}}
        \vspace{0.5\baselineskip}
    \end{figure}

An equivalent representation of Eq. (\ref{eqn:sigmod}) is to write it in its complex-valued form, using Euler's identity, by including only a single sideband of the sinusoidal term in the illumination pattern. This representation is given by
 \begin{equation}
        \Tilde{S}^{(1)}(t,\phi) = \langle \Psi_{\phi}(\vx,\Delta \vk(t)) c(\vx) \rangle_{\vx} 
        \label{eqn:forward_Model}
    \end{equation}
where $\Psi_{\phi}(\vx,\Delta \vk(t)) = \exp[i(\Delta \vk(t) \cdot \mR_\phi \vx)]$ is the complex Fourier kernel in rotated coordinates $\mR_\phi \vx$ and we have assumed $\mu(t)=1$ for simplicity. The procedure for obtaining this representation from the data is illustrated in supplementary material Fig. S2(a,b), \cite{Field2015}. The value of this complex-valued representation in Eq. (\ref{eqn:forward_Model}) is showing that each time sample corresponds to a complex amplitude of the object spatial frequency distribution --  equivalent to the data in an individual ODT  line image, establishing Eq. (\ref{eqn:forward_Model}) as the forward model for FDT, which is equivalent to the Fourier Diffraction theorem \cite{WOLF1969,DEVANEY}. We refer to Eq. (\ref{eqn:forward_Model}) as the forward model for FDT with forward operator $\mathcal{D}\{c(\vx)\}(t,\phi)=\langle \Psi_{\phi}(\vx,\Delta \vk(t)) c(\vx)\rangle_{\vx}$.

The spatial distribution of the data collected for each angle is referred to as a sinogram which is computed from a Fourier transform of the single side band time signal, $s(\mR_\phi \vx) = \mathcal{F}\{\Tilde{S}^{(1)}(t,\phi)\}$. The least-squares estimate (minimum $\mathcal{L}^2$ norm for error) of the fluorescent concentration, denoted by $\hat{c}(\vx)$, is then computed by applying the inverse operator $\mathcal{D}^{-1}$:
    \begin{equation}\label{eqn:inverse_Model}
        D^{-1} \{\Tilde{S}^{(1)}(t,\phi)\}(\vx)=\langle \Tilde{\Psi}^\dagger_{\phi}(\vx,\Delta \vk(t)) \Tilde{S}^{(1)}(t,\phi) \rangle_{t,\phi}=\hat{c}(\vx)
    \end{equation}
where $\dagger$ denotes adjoint (complex-conjugate for Fourier kernel). 
The kernel of the inverse operator, $\Tilde{\Psi}_{\phi}$, forms a biorthogonal system with the kernel of the forward operator, $\Psi_{\phi}$, at the limit of high $\mathrm{NA}$, as shown in Supplementary Material. In a noiseless case, this biorthogonality leads to a perfect reconstruction of the object.  The dual kernel is given by $\Tilde{\Psi}_{\phi} = \gamma(t) \, \Psi_{\phi}$, where $\gamma(t) =  |t|/T \sqrt{1-({\rm NA} \, t/T)^2}$ is the determinant of the Jacobian of the coordinate transformation $t \longrightarrow \Delta \vk(t)$. The estimate in Eq. (\ref{eqn:inverse_Model}) is formally equivalent to the ODT backpropagation reconstruction \cite{DEVANEY}. 

A full simulation of the forward model, Eq. (\ref{eqn:forward_Model}), and the reconstruction, Eq. (\ref{eqn:inverse_Model}), is shown in Fig. \ref{fig:reconstruction}. The FDT microscope was simulated using an illumination wavelength of $532 \text{ nm}$, $\text{NA} = 0.90$, and a field of view of $20 \, \mu \text{m}$; the full time trace signal processing workflow to generate the FDT sinogram is illustrated in Supplementary material Fig. S2. Panel (a) shows a FDT sinogram in the spatial domain where the scan angles range from $[-180,180]$ degrees. The colored lines and boxes represent the time trace(s) used to generate the corresponding figures in the right of the sinogram.  The first and third rows in Fig. \ref{fig:reconstruction} show the mapping of measured spatial frequency support onto the Ewald sphere in the spatial frequency domain constrained by optical diffraction generated by taking the Fourier transform of the second and fourth rows, respectively. Panels (b) and (d) show the frequency support measured by a time trace when $\phi = 0$ and $\phi = 45$ degrees, respectively.  Panels (f) and (h) show the spatial frequency support when multiple projections are used in the reconstruction, $\phi = [0,45]$ and $\phi = [-180,180)$ degrees, respectively. Panels (c,e,g,i) show the reconstructed object. Notice that the object localization improves as additional projection angles are used in the reconstruction.

The FDT microscope was experimentally implemented by using a spinning modulation mask that behaves as a time varying grating spatial frequency \cite{Futia:11}. The mask is illuminated by a line focus that is imaged relayed to the object region. At a snapshot in time, the mask appears as a static grating creating a zero order beam, $E_0$, as well as positive, $E_1(t)$, and negative order diffracted beams. The negative diffracted order is blocked by a spatial filter, leaving the zero and positive diffracted order beams to be image relayed to the object plane \cite{Field2015}.  Interference between the beams produces the desired spatio-temporally modulated illumination intensity pattern. The object is rotated a full 360 degrees. At each rotation angle a full time trace is acquired. Rotation in the spatial domain also causes the spatial frequency arc, Fig. \ref{fig:Scattering}(d), to rotate.  Once data from all illumination angles has been acquired, the full $(k_x-k_z)$ frequency plane will have been sampled so that we may estimate the object with isotropic spatial resolution. See the supplementary information for details on the experimental apparatus and reconstruction algorithm.

\begin{figure}[t!]
  \centering
  \fbox{\includegraphics[width=0.97\linewidth]{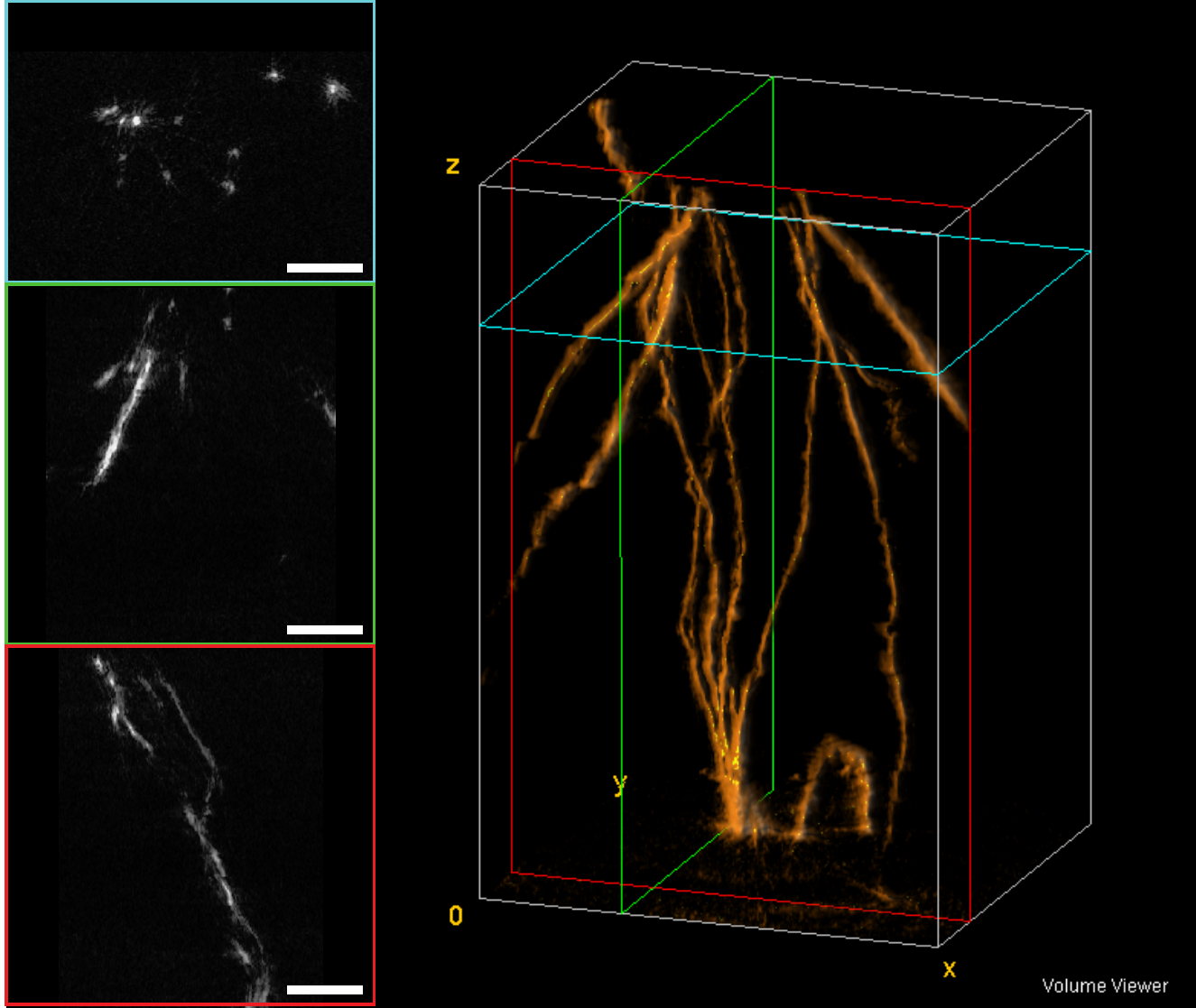}}
  \caption{3D reconstruction of fluorescent stained cotton fibers. The blue, green, and red panels are slices of the object from $x-y$, $y-z$, and $x-z$ slices, indicated by the colored rectangle in the main figure on the right. Scale bar equals $60 \mu \text{m}$.}
  \label{fig:Epi_data}
  \vspace{0.5\baselineskip}
\end{figure}

FDT imaging was demonstrated experimentally using an object fabricated from cotton fibers stained with fluorescein.  The stained fibers were mounted on a eight axis stage, Fig. S1(c). The mounting stage allowed for full 360 degree rotation of the sample as well as the ability to position the sample precisely in the microscope focus.  Fig. \ref{fig:Epi_data} shows a 3D reconstruction of fluorescein stained fibers using alpha blending from Volume Viewer in imageJ.  The image was generated with 200 evenly spaced $x-z$ slices obtained by scanning along $y$.  Each slice reconstruction used 360 measured time signals evenly spaced from $\phi = [-180, 180)$ degrees, averaged over 10 time traces.  Due to mechanical instability of the y-axis stage, each $x-z$ slice was shifted to align adjacent slices to avoid object discontinuity in the 3D reconstruction. The sub-images in Fig. \ref{fig:Epi_data} are slices from the 3D reconstruction and the colored frames correspond to the rectangular boxes in the 3D image.  An absorption contrast image was simultaneously acquired with the fluorescence, however, for brevity this image is not shown in the main text; see Fig. S3. 

There are several differences between standard optical diffraction tomography (ODT) and FDT that should be noted. While ODT and FDT obtain the complex spatial frequency values that follow the arc of spatial frequency information governed by diffraction as shown in Figs. \ref{fig:Scattering}(b) and (d), the physical origin of these data are remarkably different. ODT relies on the spatial coherence of the light scattered by the spatial variation in the refractive index of the object. As a result, ODT projection operation deviates from FDT by a complex scaling constant of $-2i \Delta k_z$. In contrast, FDT records information from the spatial variation of fluorophore concentration from the interference of two spatially coherent illumination beams. Each method samples complex amplitudes that lie on the Ewald sphere, which naturally leads to recording data in the $k_x-k_z$ spatial frequency plane by relative rotation of the object and illumination beams to spatially resolve the object in the $x-z$ plane. However, in the derivation of the Fourier diffraction theorem for FDT the only assumption made was illumination by plane waves, and there is no need to invoke the Born approximation or Rytov approximation. Therefore, FDT does not have the same object size or object variation limitations that standard ODT experiences \cite{OTD_lim}.

FDT mitigates the coupling between object size and spatial resolution typically seen in fluorescence imaging. Comparatively, in optical projection tomography, where the fluorescent light is detected with a camera, the object is restricted to the region of good focus (the Rayleigh range) to avoid background blur from out of focus light \cite{Comparison2019,fluorTom}. This causes the coupling of spatial resolution and object size conventionally seen with incoherent imaging modalities. In FDT, incoherent light emission may be treated as a coherent source allowing the object to extend over a much larger region not constrained by the Rayleigh range. Therefore, FDT decouples the need to reduce the numerical aperture of the illumination as the object size increases.

In summary, we introduced a new tomographic imaging technique, Fluorescence Diffraction Tomography (FDT), that extends optical diffraction tomography to incoherent contrast mechanisms, such as fluorescence and Raman scattering. We developed theory for both forward and inverse models.  The forward model uses CHIRPT illumination and detection as a projection of spatial frequencies onto the sample \cite{Field2015, Field:16, Stockton2017, Field:2019}. The projection uses modulation transfer to encode the spatial phase of the illumination to allow phase transfer to incoherent sources.  We demonstrate FDT reconstruction with dual functions that are biorthogonal to the intensity illumination of the rotated Fourier elements in the forward model. Additionally, we showed experimentally that FDT works for both coherent and incoherent contrast mechanisms. In principle it can be used for any contrast mechanism including nonlinear mechanisms.  We expect this technique will expand the range of samples that can be imaged and provide richer information as it is easy to co-register multiple contrast distributions simultaneously.

\section*{Funding}
We acknowledge funding support from the National Institute of Health (NIH) (R21EB025389, R21MH117786). J. Squier is supported by the National Science Foundation (NSF)(1707287).

\section*{Disclosures} The authors declare no conflicts of interest.

\section*{Supplements}

\subsection{Experimental Setup}

\begin{figure}[h!]
    \centering
    \fbox{\includegraphics[width=.7\linewidth]{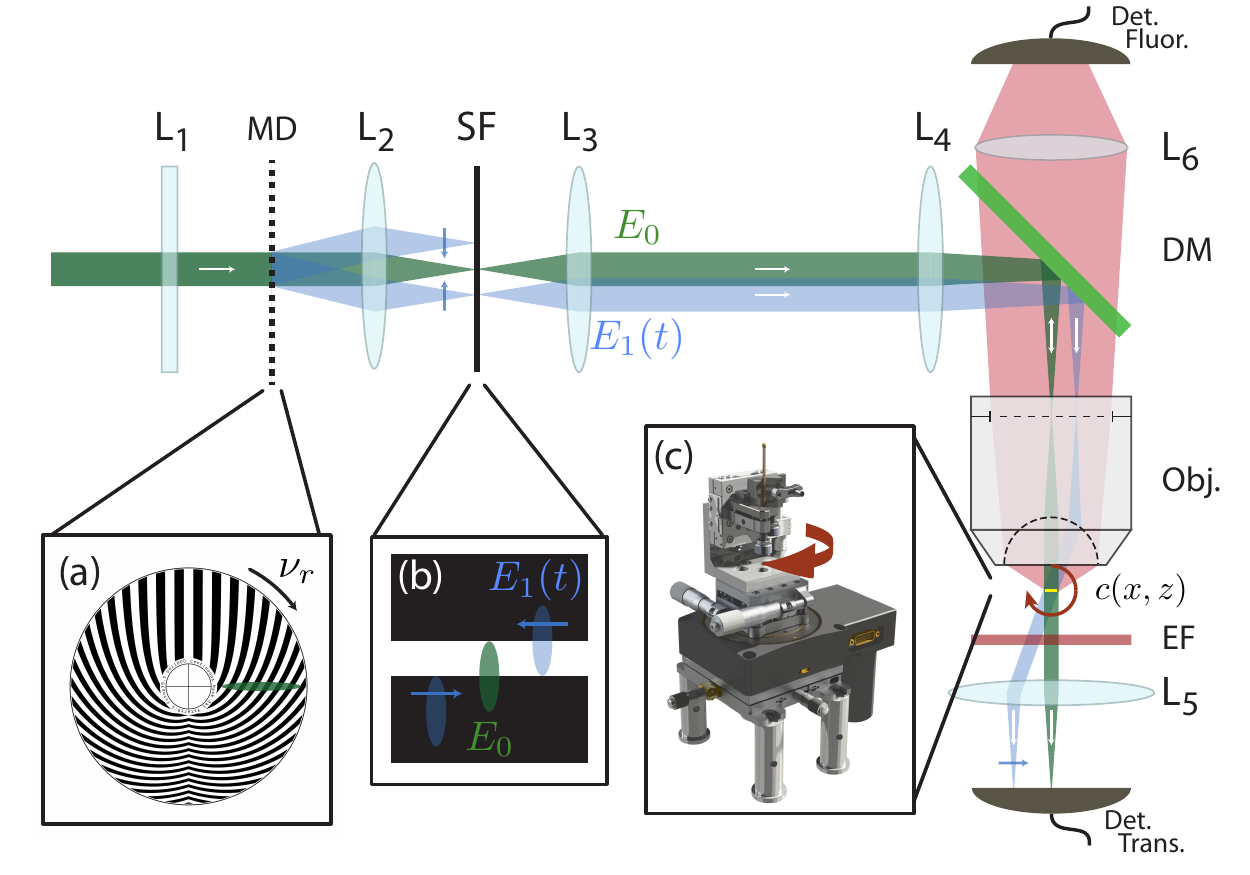}}
  \caption{Schematic of Fluorescence Diffraction Tomography microscope. Panel (a) modulator mask. Panel (b) isolating spatial filter. Panel (c) sample mounting rotation stage.  $L_1$ - Cylindrical lens, $L_{2-5}$ - Spherical lens, MD - modulator disk, SF - Spatial filter, DM - Dichroic Mirror, Obj. - Objective lens, $c(x,z)$ - Sample, EF - Emission filter, Det. - single-pixel detector, $\nu_r$ - Rotation frequency, $E_0$ - Zeroth order illumination field, $E_1(t)$ - Positive first order scan field}
  \label{fig:schematic}
\end{figure}

The experimental setup for Fluorescent Diffraction Tomography (FDT) is shown in Fig. \ref{fig:schematic}. A continuous-wave (CW) laser (Lighthouse, Sprout) wavelength, $\lambda = 532$-nm, is collimated and brought to a line focus with a cylindrical lens on a spinning modulator disk, Fig. \ref{fig:schematic}(a). The modulator is a transmission mask designed to impart a unique modulation frequency as a function of disk radius \cite{Futia:11,Field2015,Field:16,Field:18}. As the disk spins at a constant angular velocity, the transmission pattern presents a time varying grating producing diffracted orders. A slit spatial filter is placed in the back focal plane of a 2f optical system, see Fig. \ref{fig:schematic}(b), and selects only the zero and first diffracted orders \cite{Field2015} to produce a stationary reference beam and an angle scanning beam, which act as the incident field and scattered field in a coherent scattering experiment, respectively \cite{WOLF1969}. The filtered beams are image relayed to the sample region with a 4-f imaging configuration with a tube lens, $f_{tube} = 250$ mm, and objective lens, $f_{obj} = 35$ mm. The sample was mounted on a rotation stage (Newport, URB100CC) to allow full 360 degree rotation in the $x-z$ plane, Fig. \ref{fig:schematic}(c). The transmitted light was collected by a 0.25 NA aspheric lens (New Focus, 5725-A) and image relayed to a photodiode detector (Thorlabs, DET100A). The fluorescence was collected in the epi-direction and image relayed to a PMT (Hamamatsu H9305). The fluorescence was separated from the illumination light with a dichroic beamsplitter (Semrock, FF562-Di03) and an interference filter (Semrock, FF01-593/40).

The objective lens, a 35 mm focal length achromatic lens (Thorlabs, AC254-035-A), was chosen, instead of a typical high NA objective, to alleviate the transverse wobble seen by the mounting stage, ${\small \sim}30 \mu m$, which can lead to significant reconstruction distortions. In order to correct the transverse wobble, a large field of view (FOV),  ${\small \sim}260 \mu m$, was used to ensure that the sample stayed in the central region of the FOV so the sample image could be shifted laterally in post processing to remove the effect of transverse wobble.

\subsection{Mathematical description of FDT image formation}

The forward operator $\mathcal{D}$ can be viewed as a map from $\mathcal{L}^{2}(\mathbb{R}^2)$, where the concentration function $c(\vx)$ lives, to $\mathcal{L}^{2}(\mathbb{R}\times (-\pi,\pi])$, where $\Tilde{S}^{(1)}(t,\phi)$ lives:    
 \begin{equation}\label{eqn:forward_Model}
        \mathcal{D}\{c(\vx)\}= \langle \Psi_{\phi}(\vx,\Delta \vk(t)) c(\vx) \rangle_{\vx} = \Tilde{S}^{(1)}(t,\phi).
    \end{equation}
The kernel of this operator is the Fourier kernel in rotated spatial coordinates $\mR_\phi \vx$:
\begin{equation}
  \Psi_{\phi}({\vx},\Delta \vk(t)) = \exp \left(  i \,  \Delta \vk(t) \cdot \mR_\phi \vx  \right), \nonumber
\end{equation}
where
\[
\mR_{\phi}=\begin{pmatrix} \cos\phi & \sin\phi \\ -\sin\phi & \cos\phi \end{pmatrix}
\]
is a rotation matrix by $\phi\in [0,2\pi)$, and $\Delta\vk(t)=(\Delta k_x(t), \Delta k_z(t))$ is the difference wavevector parameterized by time, with $\Delta k_x(t)=k_c t/T$ and $\Delta k_z=k \, (\sqrt{1-({\rm NA} \, t/T)^2}-1)$ for $t\in [-T,T]$. Because of this parameterization, sampling in time is equivalent to sampling in spatial frequency or wavenumber on a linear grid in $\Delta k_x$ and a parabolic grid in $\Delta k_z$. 

The forward operator in \eqref{eqn:forward_Model} is an inner product in $\mathcal{L}^2(\mathbb{R}^2)$. The adjoint operator is given by  
\begin{equation}
 \mathcal{D}^{\dagger}\left\{ \Tilde{S}^{(1)} (t,\phi)\right\}({\vx}) = \int_0^{2 \pi}  \int_{-T}^{T}    \Psi^{\dagger}_{\phi}({\vx},\Delta \vk(t)) \, \tilde{S}^{(1)}(t,\phi) \, d t \,  d\phi
\end{equation}
with the kernel 
\[
\Psi^{\dagger}_{\phi}(\vx,\Delta\vk(t))= \exp \left( - i \, \Delta {\vk}(t) \cdot \mR_\phi \vx \right).
\]
Applying the adjoint operator to $\Tilde{S}^{(1)}(t)$ is equivalent to doing correlation processing (matched filtering) on the data and furnishes an estimate of $c(\vx)$. However, this estimate in general does not enjoy any sense of optimality.

The least-squares (minimum $\mathcal{L}^2$-norm error) estimate is obtained by building the inverse operator $\mathcal{D}^{-1}$: 
\begin{equation}
 \mathcal{D}^{-1}\left\{ \tilde{S}^{(1)} (t,\phi)\right\}({\vx}) = \int_0^{2 \pi}  \int_{-T}^{T}   \tilde{\Psi}^{\dagger}_{\phi}(\vx,\Delta\vk(t)) \, \tilde{S}^{(1)}(t,\phi) \,  d t \, d \phi.
 \label{eqn:inverseoperator}
\end{equation}
The kernel for $\mathcal{D}^{-1}$, called the dual kernel, is given by
\[
\Tilde{\Psi}_{\phi} (\vx,\Delta\vk(t))= \gamma(t) \, \Psi_{\phi}(\vx,\Delta\vk(t))=\gamma(t) \exp\left(i \Delta {\vk}(t) \cdot \mR_{\phi}\vx \right), 
\]
where 
\[
 \gamma (t) =  \frac{|t|/T}{\sqrt{1-({\rm NA} \, t/T)^2}}
\]
is the determinant of the Jacobian of the coordinate transformation $t \longrightarrow \Delta\vk(t)$. 

As the numerical aperture $\mathrm{NA}$ goes to $1$, this dual kernel forms a biorthogonal system with the forward kernel, and at the limit we have  
\begin{equation}\label{eqn:biorth1}
  Q:=\int_{0}^{2\pi}\int_{-T}^{T} \, \Psi^{\dagger}_{\phi}({\vx},\Delta\vk(t)) \, \tilde{\Psi}_{\phi}(\vx',\Delta\vk(t)) \, d t \, d \phi     = 
  \delta^2({\vx}-{\vx}'),
\end{equation}
where $\delta^2(\cdot)$ denotes a bivariate Dirac delta. To see this, let us define $\Delta \vx= \vx-\vx'$. Then, $Q$ can be written as 
\begin{align}
Q&=\int_{0}^{2\pi}\int_{-T}^{T} \gamma(t) \exp \left( - i \, \Delta {\vk}(t) \cdot \mR_\phi \vx \right) \exp \left( i \, \Delta {\vk}(t) \cdot \mR_\phi \vx' \right) \, dt \, d\phi\nonumber\\
&=\int_{0}^{2\pi}\int_{-T}^{T} \gamma(t) \exp \left( - i \, \Delta {\vk}(t) \cdot \mR_\phi \Delta\vx \right)\, dt \, d\phi\nonumber\\
&=2\pi\int_{-T}^{T} \gamma(t)  J_0(\|\Delta \vk(t)\|_2 \|\Delta\vx \|_2) \, dt.\nonumber
\end{align}
where $J_0(\cdot)$ is the Bessel function of the first kind in two dimensions and $\|\cdot \|_2$ denotes 2-norm. Substituting for $\Delta\vk(t)$, this can be simplified to 
\[
Q=\frac{4\pi \mathrm{NA}}{T}\int\limits_{0}^1 \frac{\tau}{\sqrt{1-(\mathrm{NA} \, \tau)^2}} \, J_0\left(\sqrt{2}  \, k \, \|\Delta \vx\|_2 \,  \sqrt{1-\sqrt{1-(\mathrm{NA}\tau)^2}}\right) \, d\tau
\]
where $\tau=t/T$. Let $\Delta \kappa(\tau) := \sqrt{2(1-\sqrt{1-(\mathrm{NA} \, \tau)^2})}$. Then, the integral $Q$ can be expressed in the simplified form

\[
Q=\frac{4\pi}{\mathrm{NA} \, T}\int\limits_{0}^{\Delta \kappa_c}  \, J_0\left(k \, \|\Delta \vx\|_2 \,  \Delta \kappa \right) \Delta \kappa \, d\Delta \kappa,
\]
where $\Delta \kappa_c = \Delta \kappa (1)= \sqrt{2 \, (1-\sqrt{1-\mathrm{NA}^2})}$. The value of the above integral is 
\[
Q=\frac{4\pi \, k \, \Delta \kappa_c}{\mathrm{NA} \, T \, \|\Delta \vx\|_2} J_1\left(k  \, \|\Delta \vx\|_2 \,  \Delta \kappa_c \right).
\]

\begin{figure}[h!]
    \centering
    \fbox{\includegraphics[width=0.75\linewidth]{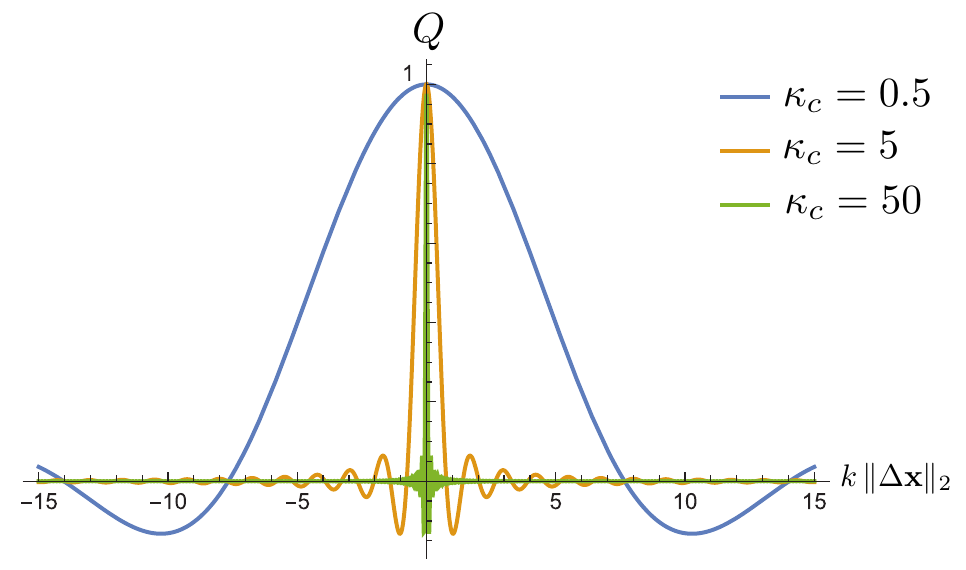}}
  \caption{Plots of the biorthogonal relationship of the relationship given in  \eqref{eqn:biorth1}. In the limit as $\kappa_c \rightarrow \inf$, the dual and forward model kernels become biorthogonal.}
  \label{fig:biorth}
\end{figure}

The biorthogonality property in \eqref{eqn:biorth1} is satisfied when $\mathrm{NA}\to 1$ or equivalently when $\Delta \kappa_c \to \sqrt{2}$, whereupon $Q \to \delta^2(\vx-\vx')$. However, physical constraints imposed by the experimental system limit the maximum value of $\Delta \kappa_c$. In the experiment, the illumination light is propagating within the region of the object and can be described by Helmholtz equation. The dispersion relationship of the Helmholtz equation requires that a plane wave propagating with the transverse spatial frequency $k_x$ carries an axial spatial frequency of $k_z = \sqrt{k^2-k^2_x}$. This condition sets the difference axial spatial frequency to $\Delta k_z = \sqrt{k^2-k^2_x} - k$. These restrictions enforce a maximum value of the difference wavevector norm of $\Delta \kappa_c = \sqrt{2}$ for $\mathrm{NA}=1$. This solution to the integral \eqref{eqn:biorth1} is plotted in Fig.~\ref{fig:biorth}.  The horizontal axis is $k \, \|\Delta\vx\|_2$. At the limiting case, and if there is no noise, this biorthogonality property results in exact reconstruction of the object through the inverse operator. For smaller numerical apertures, the width of $Q$ limits the reconstruction resolution. 

\subsection{Data Processing}

Equations (2) and (3) in the Letter describe the forward and inverse models for Fluorescence Diffraction Tomography (FDT). Here we describe the detailed signal processing steps required for processing experimental data and using that data for image reconstruction. 

\begin{figure}[h!]
    \centering
    \fbox{\includegraphics[width=0.75\linewidth]{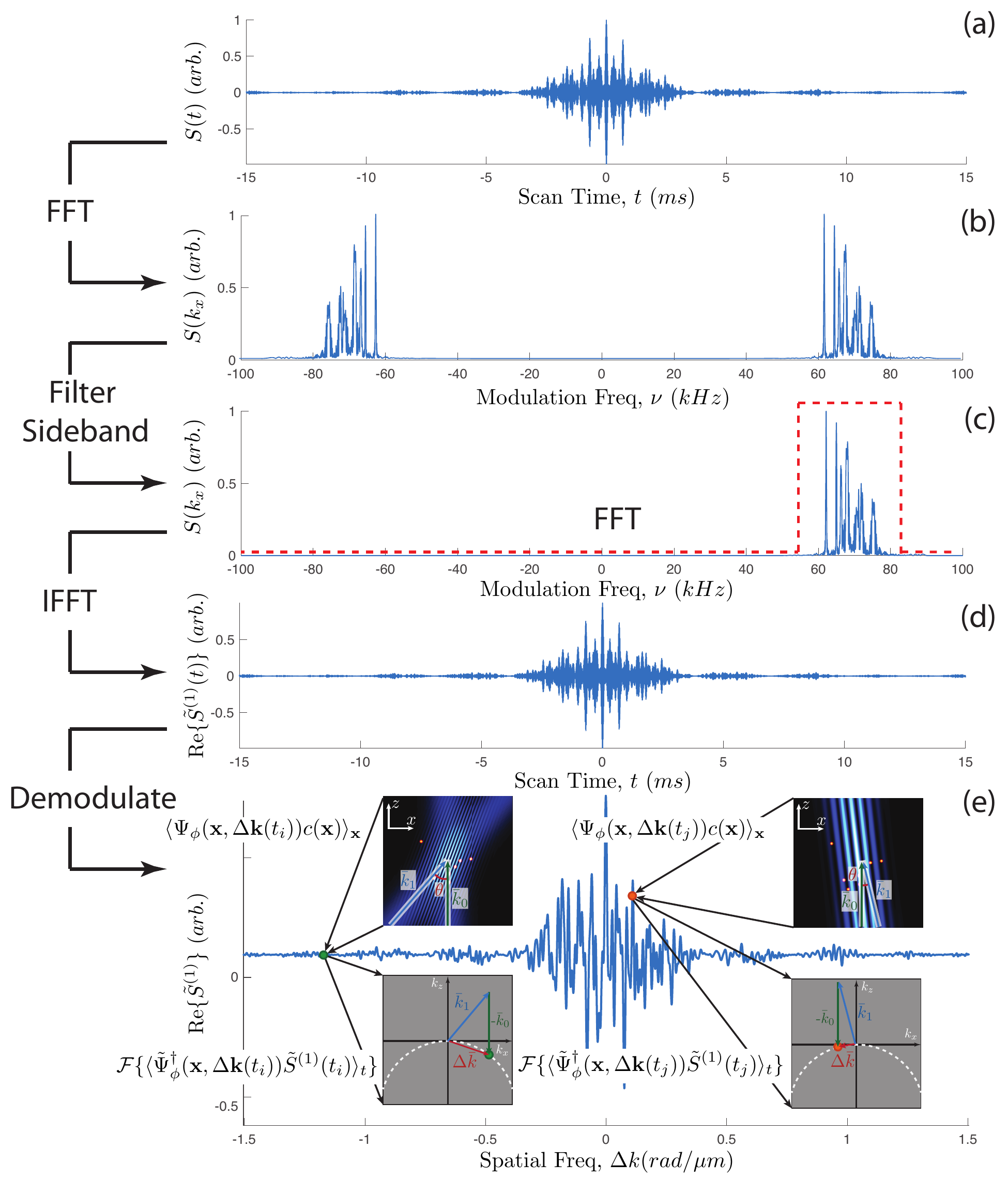}}
  \caption{The signal processing procedure to generate the FDT sinogram using simulated data. (a) A single time trace, $S(t)$, at an arbitrary angle collected by a photodiode. (b) The spectral density, $\mathcal{F}\{S(t)\}$, displays conjugate symmetry about DC modulation frequency. (c) The spectral density is filtered to isolate the positive sideband, dotted red line. (d) The inverse Fourier transform (iFFT) of the filtered spectral density encodes the spatial phase difference between the illumination beams. (e) The complex time trace is demodulated by the carrier frequency to remove linear phase ramp.  The figures above the time trace illustrate how each time point is generated by two beam interference and spatial integration by the single-pixel detector, $x-z$ space.  The figures below the time trace illustrate how each measured time point represents the complex object spatial frequency, which is mapped to the Ewald sphere in the FDT reconstruction, $k_x-k_z$ space.}
  \label{fig:SignalFlow}
\end{figure}

As a single scan is taken over one rotation of the modulator mask, the time trace, modeled by Eq.~(1), is generated by collecting the signal light on a single-pixel photodetector  (Fig.~\ref{fig:SignalFlow}(a)). The complex demodulated sideband given in Eq.~(2) is isolated by first taking a simple Fourier transform of the time trace.  The carrier frequency, $\omega_c$, causes the spatial distribution of the projection, $s_{\phi}(x_{\phi})$, to be centered at $+\omega_c$ and a conjugate image is centered at $-\omega_c$, as shown in Fig.~\ref{fig:SignalFlow}(b) \cite{Field2015,Field2016PNAS}. The carrier frequency plays a role that is analogous to the off-axis reference beam for holography, avoiding the twin image problem \cite{Leith:62}. 

In order to recover the complex object information, the positive single side band is isolated by applying a bandpass filter in the frequency domain, shown as a red dotted line in \ref{fig:SignalFlow}(c). Once the bandpass filter has been applied, the signal is converted back into the time domain by taking the inverse Fourier transform. This operation results in a complex time signal that contains the carrier frequency, i.e., the rapid oscillation of the real part of the time signal shown in Fig. \ref{fig:SignalFlow}(d). This complex temporal data is then demodulated by the carrier frequency  to bring the line image information to the base band (Fig.~\ref{fig:SignalFlow}(e)). In total, these operations provide the complex single sideband given by Eq.~(2). At this point in the reconstruction workflow, known optical aberrations can be corrected, such as spherical aberration and so-called wobble phase imparted by an imperfectly mounted disc \cite{Field2015, aber_correction, Field:16, Field2016PNAS}. 

The demodulated time signal is downsampled to reduce the data size and speed up the reconstruction algorithm, thereby and reducing data pressure. Since the time signal is already band limited, it is not necessary to apply a lowpass filter in the downsampling operation. The figure in Fig.~\ref{fig:SignalFlow}(e) shows the real part of the demodulated single sideband signal, $\tilde{S}^{(1)}(t)$, which is used in the FDT reconstruction algorithm (main text, Eq.~(3)).  Scanning the object over $\theta \in (-180,180]$ degrees, processing the time traces as described above, and Fourier transforming the downsampled, demodulated, single side band, time trace results in the sinogram shown in the main text, Fig.~2(a).

Each time point in $\Tilde{S}^{(1)}(t)$ is a measurement of the magnitude and phase of object spatial frequency representation at the instantaneous projected spatial frequency pair $(\Delta k_x(t), \Delta k_z(t))$ \cite{Field2015, Field:16, Field2016PNAS}. The inlaid figures in Fig. \ref{fig:SignalFlow}(e), above the time trace, show the illumination intensity at a snapshot in time, which is the interference between the reference beam, $E_0$, and the scanning beam, $E_1(t)$, at a crossing angle, $\theta(t)$, which produces a spatial frequency $\Delta \vk(t)$. The illumination intensity excites the fluorescent concentration distribution at the given spatial frequency. The emitted fluorescent light is collected by a single-pixel detector, performing a spatial integration along the spatial coordinates $x$ and $z$, modeled by Eq. (1). Note that in this work, we assume the detector is infinite in extent for simplicity in our forward model. This large detector size is a good approximation to the experimental system that we have described here; although it should be noted that there are cases in which the finite size of the detector and point-spread-function of detection must be accounted for \cite{Field:2019}. 

The measured spatial frequency information is mapped into the object spatial domain by the inverse operator \eqref{eqn:inverseoperator}. The action of the inverse operator is illustrated by the inlaid figures below the time trace in Fig.~\ref{fig:SignalFlow}(e). These inlays illustrate how the measured spatial frequency information is mapped into the object spatial distribution with  \eqref{eqn:inverseoperator}.  As the disc rotates, each measured spatial frequency component is obtained from an arc that traverses the Ewald sphere over the range of transverse spatial frequencies supported by the NA of the illumination objective. In this way all spatial frequencies supported by the illumination objective are sequentially scanned as the field, $E_1$, passes through the full spatial frequency support of the imaging system \cite{Field2015, Field:18}. The green and orange points illustrate high and low spatial frequencies, respectively.

Due to the fact that we sample discrete time points, it is appropriate to express the inverse operator, \eqref{eqn:inverseoperator}, as a Riemann sum. The time points, at time $t_i$, are samples on a regularly spaced grid with a time step $\Delta t$. A fixed set of rotation angles, $\phi_j$, are acquired over an angular span of $2 \, \pi$ with uniform angular spacing $\Delta \phi$, using the index $j$. This discrete representation leads us to rewrite Eq.~(3) in the main text as
\begin{equation}
    \hat{c}(x,z) = \sum_{i=0}^{T} \sum_{j=1}^{2 \, \pi} \tilde{\Psi}_{i,j} \tilde{S}_{i,j}^{(1)} 
    \label{eqn:discreteRecon}
\end{equation}
which can be written explicitly as 
\begin{equation}
    \hat{c}(x,z)  = \sum_{i=0}^{T} \sum_{j=1}^{360} \gamma_i \mathrm{exp}\big[i(\Delta  k_{x_i} x_{\phi_j} + \Delta k_{z_i} z_{\phi_j})\big] \tilde{S}_{i,j}^{(1)}
    \label{eqn:discreteRecon}
\end{equation}
\noindent where $\gamma_i = k|\Delta k_{x_i}|/(\sqrt{k^2 - [\Delta k_{x_i}]^2})$ is the magnitude of the determinant of the Jacobian, and $\Delta k_{x_i}$ and $\Delta k_{z_i}$ are difference wavenumbers in $x$ and $z$, respectively, $x_{\phi_j}$ is the rotated $x$-coordinate vector, and $z_{\phi_j}$ is the $z$-coordinate vector. From \eqref{eqn:discreteRecon}, we see that the reconstruction algorithm applied to discrete data is performed in the spatial domain by weighting the dual operator kernel functions, that are sampled on the discrete reconstruction spatial grid, by the complex, demodulated, single sideband time signal samples.

\subsection{Absorption Contrast}

\begin{figure}[h!]
\centering
\fbox{\includegraphics[width=0.9\linewidth]{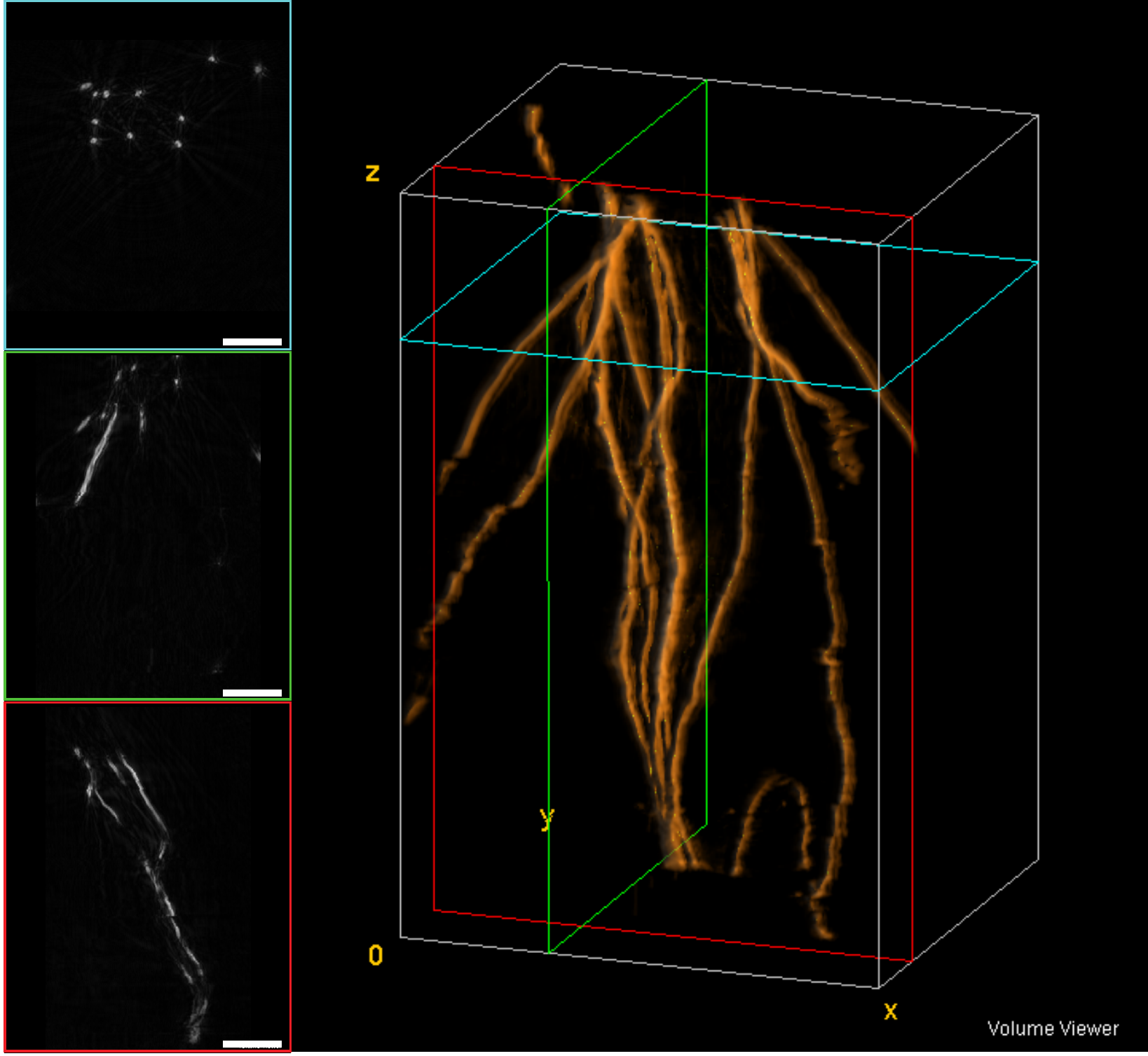}}
  \caption{3D reconstruction of absorption contrast of fluorescent stained cotton fibers. The blue panel is an $x-y$ slice of the object indicated by the blue rectangle in the main figure on the right. The green panel is an $y-z$ slice indicted by the green rectangle. The red panel is an $x-z$ slice indicated by the red rectangle. Scale bar equal $60 \mu \text{m}$.}
  \label{fig:Trans_data}
\end{figure}

FDT can be used for many contrast mechanisms (we demonstrate fluorescence and absorption), regardless of the whether the light emerging from the specimen is coherent or incoherent. While in the main text we have primarily focused on FDT for fluorescent light, we also demonstrate the ability to extend fluorescent diffraction tomography to simultaneously image light lost to absorption or scattering in transmission. Here, we demonstrate that absorption contrast mechanism, which was simultaneously acquired along with fluorescence on a separate single pixel detector. The absorption signal was acquired by collecting the transmitted excitation light, $532\ \mathrm{nm}$, on a photodiode (Thorlabs, DET100A). Fig. \ref{fig:Trans_data} shows the 3D reconstruction of the absorption contrast. The left hand columns shows slices along the $x-y$, $y-z$, and $x-z$ planes in descending order, respectively.

\subsection{FDT Comparison with Backprojection}

In the introduction of the main text, we mention one of the major limitations of Optical Projection Tomography (OPT) and Selective Plane Illumination Microscopy (SPIM) is the fact that both techniques exhibit coupling between the object size and spatial resolution \cite{DiffTomMicr,fluorTom,spim,Comparison2019}. That is, the object being imaged may not be larger than roughly twice the Rayleigh range of the line focus \cite{Comparison2019}. The reason for this limitation is that both OPT and SPIM assume that the illumination light is approximately planar in the object region so that diffraction is negligible. These assumptions are only valid inside the Rayleigh range of a beam focus (or the focal region of the point spread function). Therefore, larger objects require a thicker line focus of illumination for SPIM to extend the Rayleigh range over the object, and in the case of OPT, a lower NA objective is used so that the PSF is approximately collimated throughout the object thickness \cite{Comparison2019}. If this assumption is violated, an out of focus blur will result in the final reconstructed image. It is worth noting that several SPIM approaches have appeared that use virtual light sheets with PSF engineering approaches to create diffraction-free light sheets to circumvent this coupling \cite{Planchon2011}. 

With FDT, we do not make an assumption of planar illumination since the propagation phase is directly encoded in the measured temporal data. This allows FDT to numerically refocus the entire volume measured by the illumination beams in a similar manner to holography. In previous work we have demonstrated that the technique underlying FDT, CHIRPT microscopy, not only enables this holographic refocusing of fluorescent images, but also exhibits a depth-of-field up to 83$\times$ that of conventional imaging at the same NA without sacrificing spatial resolution ($\sim$440~$\mu$m) \cite{Field:16}. This DOF was measured to extend beyond 1~cm with the same optical components, but at the cost of a loss of spatial frequency support.  In the reconstruction presented here, it is not necessary to explicitly backpropagate or numerically refocus the line image \cite{Field:16}. Instead, the reconstruction algorithm uses  dual operator to directly synthesise the object in $x-z$. The sum over the kernel function of the inverse dual operator weighted by the complex demodulated values of the time trace recorded for each rotation angle $\phi$ directly gives the backpropagated object distribution for that measurement angle. The encoded propagation phase allows objects to reside well outside the Rayleigh range of the focus of the illumination light sheet.  This allows FDT to effectively decouple the spatial resolution from the maximum object size.

\begin{figure}[h!]
    \centering
    \fbox{\includegraphics[width=0.9\linewidth]{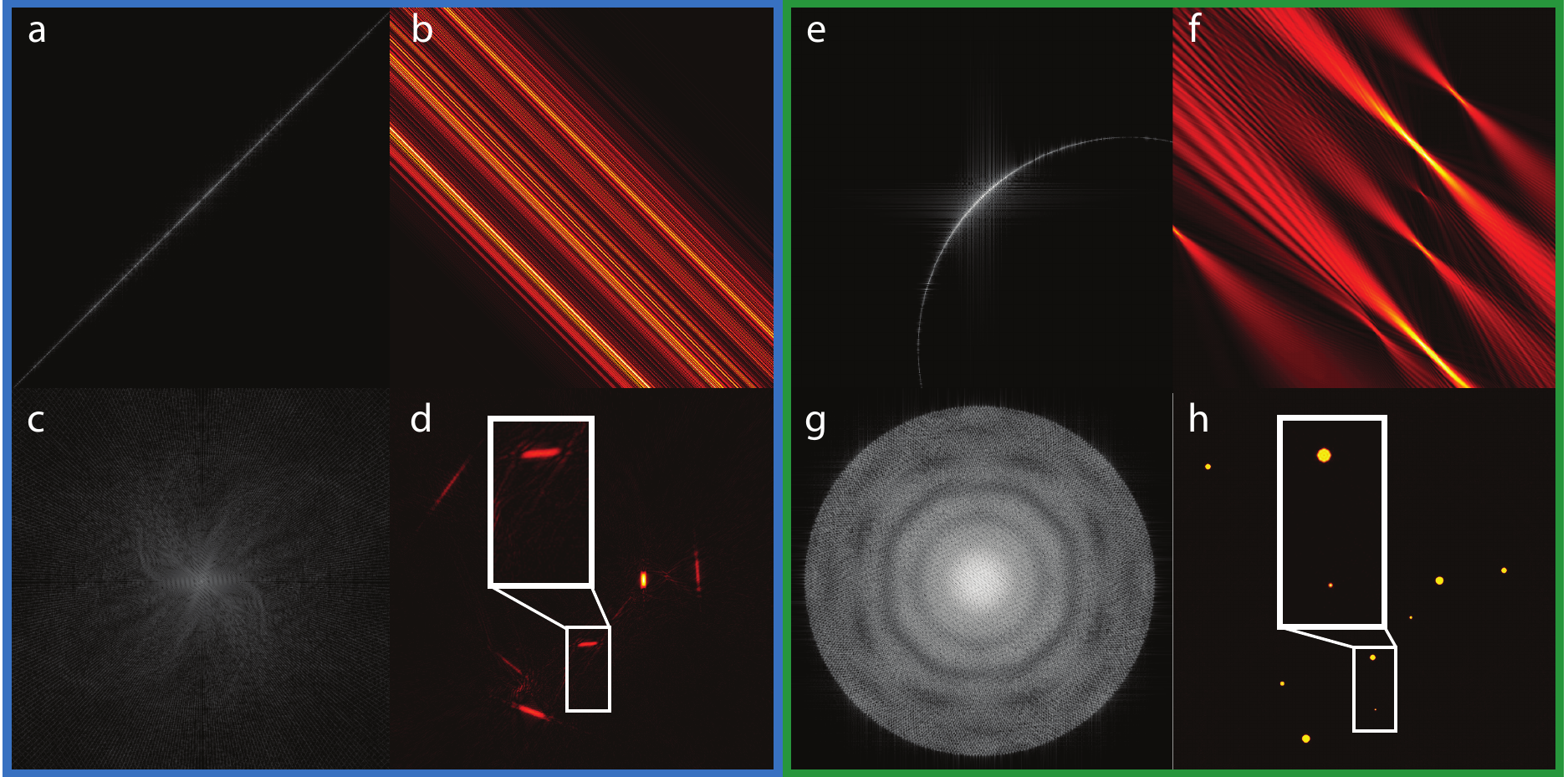}}
  \caption{Comparison between computed tomography using fluorescence intensity and Fluorescence Diffraction Tomography with phase encoding. Panel (a) shows the 2D spatial frequency support of computed tomography measured at a 45 degree angle. Panel (b) is the 2D FFT of (a) resulting in a 2D image of the object. Panel (c) shows the frequency support when the full sinogram is used in the computed tomography reconstruction. Panel (d) shows the 2D reconstructed object when diffraction is not accounted for in the reconstruction. Panel (e) illustrates how FDT, using modulation transfer, can extract the complex phase information, even from fluorescence, to map the measured spatial frequencies onto the Ewald sphere. Panel (f) shows the 2D object reconstruction using only one line image at a 45 degree angle. Panel (e) shows the full spatial frequency support when the full sinogram is used in the FDT reconstruction. Panel (h) shows the 2D reconstructed object generated by taking the 2D FFT of panel (g).}
  \label{fig:iRadon}
\end{figure}

In order to illustrate the importance of encoding the propagation phase has on the reconstruction and the effect it has on the maximum aberration free field-of-view, we perform two simulations. We start by simulating a sinogram using fluorescence as the contrast mechanism. The sinogram was generated  using an illumination wavelength, $\lambda = 532\ \mathrm{nm}$ , a numerical aperture, $\mathrm{NA} = 0.90$, field of view, $\mathrm{FOV} = 20\ \mu \mathrm{m}$, and 360 evenly spaced illumination angles ranging from $\theta = [-180,180)\ \mathrm{deg}$, sinogram shown in the main text, Fig. 2(a). 

Using the sinogram, two reconstruction strategies were tested. In the first, we reconstructed the object without the phase information that we obtained from the complex spatial frequency values by using the magnitude of the sinogram, which is equivalent to the information that would be obtained in OPT if the object size was much larger than the Rayleigh range. We reconstructed this data with the filtered backprojection algorithm using an inverse Radon transform  ($\mathrm{iradon()}$ function in Matlab). The inverse Radon transform is relevant for line-projection tomographies \cite{LT,FCT}. The results of this reconstruction strategy are shown in Figs. \ref{fig:iRadon}(a-d). Fig. \ref{fig:iRadon} (a) is the 2D frequency support measured at $\theta = 45\ \mathrm{deg}$. Notice, the frequency support lies on a straight line in the $(k_x-k_z)$ plane at a 45 degree angle, corresponding to the illumination angle, in agreement with the Fourier slice theorem \cite{Kak2001}. Fig. \ref{fig:iRadon}(b) is the resulting 2D object reconstruction in the $x-z$ plane, generated by taking the 2D inverse Fourier transform of Fig. \ref{fig:iRadon}(a).  The reconstructed object does not exhibit diffraction, i.e., the objects appear as uniform lines of constant magnitude rotated by $45^{\circ}$, which is in accordance with the assumptions made by OPT.  Figure~\ref{fig:iRadon}(c) is the frequency support using the full sinogram. The resulting 2D $x-z$ object reconstruction, Fig. \ref{fig:iRadon}(d), was generated by taking the 2D inverse Fourier transform of Fig. \ref{fig:iRadon}(c).  A radially dependent azimuthal blurring is evident in the reconstructed object. This result is expected since the Rayleigh range of the simulated illumination beam was $\small\sim 0.365\ \mu \mathrm{m}$, while the reconstructed field of view was $20\ \mu\mathrm{m}$. 

By contrast, we compare these same conditions to the inverse operator using Eq. (3) from the Letter. Figures \ref{fig:iRadon}(e-h) show those results. The reconstructed object using one angle of the discrete algorithm in \eqref{eqn:discreteRecon}, here at  $\theta = 45\ \mathrm{deg}$, is shown in  Fig. \ref{fig:iRadon}(f). Taking a two dimensional FFT of this reconstructed object gives the the measured spatial frequency support for the single illumination angle of $\theta = 45\ \mathrm{deg}$ that is shown in Fig. \ref{fig:iRadon}(e). Notice that the frequency support lines on a curve that is predicted by Fourier diffraction theory.  The full object reconstruction using a full $2 \pi$ angles is shown in  Fig. \ref{fig:iRadon}(h). The spatial frequency support obtained from the 2D FFT of the object estimate is shown in Fig. \ref{fig:iRadon}(g). We see that the reconstructed object does not contain an azimuthal blur and the points localize the the object position, only limited by the diffraction limit set by the NA of the objective.

We note that this simulation considered planar illumination in the $x-z$-plane only, neglecting effects caused by the DOF of the light sheets used for illumination in the orthogonal ($y$) dimension. We have previously described that CHIRPT has two distinct DOF metrics -- one based on intensity modulations in the $x-z$ plane, and the other dictated by the properties of the focused illuminating light sheet \cite{Field:2019}. While this DOF in the vertical dimension would ultimately place a practical limit on the spatial resolution in $y$, the fidelity of the reconstructed image in the $x-z$-plane should not suffer. Further, we note that a variety of PSF engineering strategies could be employed to maintain spatial resolution in the vertical dimension over the full FOV \cite{Field:2019}.

\printbibliography

\end{document}